\documentclass[12pt]{article}
\usepackage{epsfig,amsfonts,amssymb}
\usepackage{hyperref}
\usepackage{comment}
\usepackage{cite}
\input epsf.sty
\usepackage{cite}

\def\ZZZ{{\hbox{ Z\kern-1.6mm Z}}}
\def\RRR{{\hbox{ R\kern-2.4mm R}}}
\def\CCC{{\hbox{ C\kern-2.5mm C}}}
\def\zzz{{\hbox{z\kern-1mm z}}}

\newcommand{\qeq}{{\hbox{=\kern-2.3mm ? \kern.5mm }}}
\renewcommand{\qeq}{=}

\newcommand{\VV}{{\cal V}}

\newcommand{\WW}{{\cal W}}
\newcommand{\BBB}{{\cal B}}

\newcommand{\AAA}{{\cal A}}

\newcommand{\MM}{{\cal M}}
\newcommand{\CC}{{\cal C}}

\newcommand{\be}{\begin{equation}}
\newcommand{\ee}{\end{equation}}
\newcommand{\ben}{\begin{eqnarray}\displaystyle}
\newcommand{\een}{\end{eqnarray}}

\newcommand{\refb}[1]{(\ref{#1})}
\newcommand{\p}{\partial}
\newcommand{\sectiono}[1]{\section{#1}\setcounter{equation}{0}}

\def\one{{\hbox{ 1\kern-.8mm l}}}
\def\zero{{\hbox{ 0\kern-1.5mm 0}}}

\newcommand{\bea}[1]{\begin{eqnarray}\label{#1} }
\newcommand{\eea}{\end{eqnarray}}

\newcommand{\eqref}{\refb}

\usepackage{bm}
\usepackage[table]{xcolor}

\def\cdnote#1{{\color{green} #1}} \def\cdnote#1{{\color{black} #1}}

\newcommand{\non}{\nonumber}

\def\Imm{{\rm Im\, }}

\def\Rea{{\rm Re\, }}

\def\figone{

\def\JPicScale{0.6}
\ifx\JPicScale\undefined\def\JPicScale{1}\fi
\unitlength \JPicScale mm
\begin{picture}(220,95)(0,0)
\linethickness{0.3mm}
\put(52.5,52.5){\circle{7.07}}

\linethickness{0.3mm}
\put(110,68){\circle{7.07}}

\linethickness{0.3mm}
\put(85,38){\circle{7.07}}

\linethickness{0.3mm}
\put(110,22){\circle{7.07}}

\linethickness{0.3mm}
\put(110,50){\circle{7.07}}

\linethickness{0.3mm}
\put(170,50){\circle{7.07}}

\linethickness{0.3mm}
\multiput(20,70)(0.24,-0.12){125}{\line(1,0){0.24}}
\linethickness{0.3mm}
\multiput(20,30)(0.18,0.12){167}{\line(1,0){0.18}}
\linethickness{0.3mm}
\multiput(55,55)(0.6,0.12){87}{\line(1,0){0.6}}
\linethickness{0.3mm}
\multiput(75,90)(0.18,-0.12){174}{\line(1,0){0.18}}
\linethickness{0.3mm}
\multiput(90,92)(0.12,-0.12){167}{\line(1,0){0.12}}
\linethickness{0.3mm}
\put(110,53){\line(0,1){11}}
\linethickness{0.3mm}
\multiput(88,38)(0.18,0.12){104}{\line(1,0){0.18}}
\linethickness{0.3mm}
\multiput(84,35)(0.24,-0.12){98}{\line(1,0){0.24}}
\linethickness{0.3mm}
\multiput(55,50)(0.3,-0.12){90}{\line(1,0){0.3}}
\linethickness{0.3mm}
\multiput(113,23)(0.24,0.11){228}{\line(1,0){0.24}}
\linethickness{0.3mm}
\put(113,50){\line(1,0){53}}
\linethickness{0.3mm}
\multiput(114,68)(0.36,-0.12){148}{\line(1,0){0.36}}
\linethickness{0.3mm}
\qbezier(55,55)(75.83,54.97)(89.06,57.38)
\qbezier(89.06,57.38)(102.3,59.78)(110,65)
\linethickness{0.3mm}
\multiput(173,50)(0.24,0.12){83}{\line(1,0){0.24}}
\linethickness{0.3mm}
\multiput(172,47)(1.19,-0.12){32}{\line(1,0){1.19}}
\linethickness{0.3mm}
\multiput(77,10)(0.36,0.12){83}{\line(1,0){0.36}}

\put(20,50){$P_{(\alpha_1)}$}

\put(75,95){$P_{(\alpha_2)}$}

\put(60,8){$P_{(\alpha_3)}$}

\put(190,50){$P_{(\alpha_4)}$}

\end{picture}

}

\def\figtwo{

\def\JPicScale{0.8}
\ifx\JPicScale\undefined\def\JPicScale{1}\fi
\unitlength \JPicScale mm
\begin{picture}(123.54,73.54)(0,0)
\linethickness{0.3mm}
\put(32,50){\circle{7.07}}

\linethickness{0.3mm}
\put(120,50){\circle{7.07}}

\linethickness{0.3mm}
\put(70,68){\circle{7.07}}

\linethickness{0.3mm}
\put(90,50){\circle{7.07}}

\linethickness{0.3mm}
\put(70,32){\circle{7.07}}

\linethickness{0.3mm}
\multiput(35,50)(0.18,0.10){175}{\line(1,0){0.18}}
\linethickness{0.3mm}
\multiput(35,50)(0.18,-0.10){175}{\line(1,0){0.18}}
\linethickness{0.3mm}
\put(70,35){\line(0,1){30}}
\linethickness{0.3mm}
\multiput(73,68)(0.36,-0.12){125}{\line(1,0){0.36}}
\linethickness{0.3mm}
\multiput(72,65)(0.18,-0.12){93}{\line(1,0){0.18}}
\linethickness{0.3mm}
\multiput(73,32)(0.36,0.12){125}{\line(1,0){0.36}}
\linethickness{0.3mm}
\multiput(70,35)(0.24,0.135){83}{\line(1,0){0.24}}
\linethickness{0.3mm}
\put(93,50){\line(1,0){24}}

\linethickness{0.3mm}
\multiput(15,65)(0.18,-0.12){93}{\line(1,0){0.18}}
\linethickness{0.3mm}
\multiput(13,35)(0.24,0.135){83}{\line(1,0){0.24}}

\linethickness{0.3mm}
\multiput(123,49)(0.18,-0.12){93}{\line(1,0){0.18}}
\linethickness{0.3mm}
\multiput(123,51)(0.24,0.135){83}{\line(1,0){0.24}}

\put(130,50){$-P_{(\beta)}$}

\put(10,50){$P_{(\beta)}$}

\end{picture}

}

\def\figreduced{

\def\JPicScale{0.8}
\ifx\JPicScale\undefined\def\JPicScale{1}\fi
\unitlength \JPicScale mm
\begin{picture}(120,75)(0,0)
\linethickness{0.3mm}
\put(20,20){\line(0,1){40}}
\linethickness{0.3mm}
\linethickness{0.3mm}
\put(20,20){\line(1,0){40}}
\linethickness{0.3mm}
\put(60,20){\line(0,1){40}}
\linethickness{0.3mm}
\put(20,60){\line(1,0){40}}
\linethickness{0.3mm}
\put(60,60){\line(1,0){40}}
\linethickness{0.3mm}
\put(100,20){\line(0,1){40}}
\linethickness{0.3mm}
\linethickness{0.3mm}
\put(60,20){\line(1,0){40}}
\linethickness{0.3mm}
\multiput(20,60)(0.12,-0.12){333}{\line(1,0){0.12}}
\linethickness{0.3mm}
\linethickness{0.3mm}
\multiput(60,20)(0.12,0.12){333}{\line(1,0){0.12}}
\linethickness{0.3mm}
\multiput(0,70)(0.24,-0.12){83}{\line(1,0){0.24}}
\linethickness{0.3mm}
\multiput(10,75)(0.12,-0.18){83}{\line(0,-1){0.18}}
\linethickness{0.3mm}
\multiput(0,10)(0.24,0.12){83}{\line(1,0){0.24}}
\linethickness{0.3mm}
\multiput(10,5)(0.12,0.18){83}{\line(0,1){0.18}}
\linethickness{0.3mm}
\multiput(100,60)(0.24,0.12){83}{\line(1,0){0.24}}
\linethickness{0.3mm}
\multiput(100,60)(0.12,0.18){83}{\line(0,1){0.18}}
\linethickness{0.3mm}
\multiput(100,20)(0.24,-0.12){83}{\line(1,0){0.24}}
\linethickness{0.3mm}
\multiput(100,20)(0.12,-0.18){83}{\line(0,-1){0.18}}
\put(5,75){\makebox(0,0)[cc]{$P_{(\alpha_1)}$}}

\put(0,5){\makebox(0,0)[cc]{$P_{(\alpha_2)}$}}

\put(115,75){\makebox(0,0)[cc]{$P_{(\alpha_3)}$}}

\put(115,5){\makebox(0,0)[cc]{$P_{(\alpha_4)}$}}

\put(40,59.8){\makebox(0,0)[cc]{$\rightarrow$}}

\put(40,40){\makebox(0,0)[cc]{$\nwarrow$}}

\put(20,40){\makebox(0,0)[cc]{$\downarrow$}}

\put(35,20){\makebox(0,0)[cc]{$\rightarrow$}}

\put(60,40){\makebox(0,0)[cc]{$\uparrow$}}

\put(80,40){\makebox(0,0)[cc]{$\nearrow$}}

\put(80,59.8){\makebox(0,0)[cc]{$\rightarrow$}}

\put(80,20){\makebox(0,0)[cc]{$\rightarrow$}}

\put(100,40){\makebox(0,0)[cc]{$\downarrow$}}

\end{picture}

}

\def\figarrow{

\def\JPicScale{0.8}
\ifx\JPicScale\undefined\def\JPicScale{1}\fi
\unitlength \JPicScale mm
\begin{picture}(110,70)(0,0)
\linethickness{0.6mm}
\multiput(20,60)(0.12,-0.12){167}{\line(1,0){0.12}}
\linethickness{0.6mm}
\multiput(20,20)(0.12,0.12){167}{\line(1,0){0.12}}
\linethickness{0.6mm}
\put(20,60){\line(1,0){40}}
\linethickness{0.6mm}
\multiput(40,40)(0.12,0.12){167}{\line(1,0){0.12}}
\linethickness{0.6mm}
\put(20,20){\line(1,0){80}}
\linethickness{0.6mm}
\multiput(60,60)(0.12,-0.12){333}{\line(1,0){0.12}}
\linethickness{0.6mm}
\multiput(10,70)(0.12,-0.12){83}{\line(1,0){0.12}}
\linethickness{0.6mm}
\put(10,60){\line(1,0){10}}
\linethickness{0.6mm}
\put(10,20){\line(1,0){10}}
\linethickness{0.6mm}
\multiput(10,10)(0.12,0.12){83}{\line(1,0){0.12}}
\linethickness{0.6mm}
\multiput(60,60)(0.12,0.12){83}{\line(1,0){0.12}}
\linethickness{0.6mm}
\put(60,60){\line(1,0){10}}
\linethickness{0.6mm}
\put(100,20){\line(1,0){10}}
\linethickness{0.6mm}
\multiput(100,20)(0.12,-0.12){83}{\line(1,0){0.12}}
\put(35,60){\makebox(0,0)[cc]{$\rightarrow$}}

\put(30,50){\makebox(0,0)[cc]{$\searrow$}}

\put(30,30){\makebox(0,0)[cc]{$\nearrow$}}

\put(50,50){\makebox(0,0)[cc]{$\nearrow$}}

\put(60,20){\makebox(0,0)[cc]{$\rightarrow$}}

\put(80,40){\makebox(0,0)[cc]{$\searrow$}}

\linethickness{0.1mm}
\put(50,10){\line(0,1){70}}

\end{picture}

}

\def\figarrowred{

\def\JPicScale{0.8}
\ifx\JPicScale\undefined\def\JPicScale{1}\fi
\unitlength \JPicScale mm
\begin{picture}(110,70)(0,0)
\linethickness{0.6mm}
\multiput(20,60)(0.12,-0.12){167}{\line(1,0){0.12}}
\linethickness{0.6mm}
\multiput(20,20)(0.12,0.12){167}{\line(1,0){0.12}}
\linethickness{0.6mm}
\put(20,60){\line(1,0){40}}
\linethickness{0.6mm}
\multiput(40,40)(0.12,0.12){167}{\line(1,0){0.12}}
\linethickness{0.6mm}
\put(20,20){\line(1,0){80}}
\linethickness{0.6mm}
\multiput(60,60)(0.12,-0.12){333}{\line(1,0){0.12}}
\linethickness{0.6mm}
\multiput(10,70)(0.12,-0.12){83}{\line(1,0){0.12}}
\linethickness{0.6mm}
\put(10,60){\line(1,0){10}}
\linethickness{0.6mm}
\put(10,20){\line(1,0){10}}
\linethickness{0.6mm}
\multiput(10,10)(0.12,0.12){83}{\line(1,0){0.12}}
\linethickness{0.6mm}
\multiput(60,60)(0.12,0.12){83}{\line(1,0){0.12}}
\linethickness{0.6mm}
\put(60,60){\line(1,0){10}}
\linethickness{0.6mm}
\put(100,20){\line(1,0){10}}
\linethickness{0.6mm}
\multiput(100,20)(0.12,-0.12){83}{\line(1,0){0.12}}
\put(35,60){\makebox(0,0)[cc]{{$\rightarrow$}}}
\put(36.5,60){\makebox(0,0)[cc]{{$\rightarrow$}}}

\put(30,50){\makebox(0,0)[cc]{{$\searrow$}}}
\put(31,49){\makebox(0,0)[cc]{{$\searrow$}}}

\put(30,30){\makebox(0,0)[cc]{{$\nearrow$}}}
\put(31,31){\makebox(0,0)[cc]{{$\nearrow$}}}

\put(50,50){\makebox(0,0)[cc]{{$\nearrow$}}}
\put(51,51){\makebox(0,0)[cc]{{$\nearrow$}}}

\put(60,20){\makebox(0,0)[cc]{{$\rightarrow$}}}
\put(61.5,20){\makebox(0,0)[cc]{{$\rightarrow$}}}

\put(80,40){\makebox(0,0)[cc]{{$\searrow$}}}
\put(81,39){\makebox(0,0)[cc]{{$\searrow$}}}

\linethickness{0.1mm}
\put(50,10){\line(0,1){70}}

\end{picture}

}

\def\figzero{

\def\JPicScale{0.6}
\ifx\JPicScale\undefined\def\JPicScale{1}\fi
\unitlength \JPicScale mm
\begin{picture}(100,75)(0,0)
\linethickness{0.3mm}
\multiput(30,70)(0.12,-0.12){500}{\line(1,0){0.12}}
\linethickness{0.3mm}
\multiput(30,10)(0.12,0.12){500}{\line(1,0){0.12}}
\put(60,60){\makebox(0,0)[cc]{$V^+$}}

\put(60,20){\makebox(0,0)[cc]{$V^-$}}

\put(85,40){\makebox(0,0)[cc]{$W^+$}}

\put(30,40){\makebox(0,0)[cc]{$W^-$}}

\put(53,74){\makebox(0,0)[cc]{$q^0$}}

\put(60,73){\makebox(0,0)[cc]{$\uparrow$}}

\put(110,40){\makebox(0,0)[cc]{$q^1 \, \rightarrow$}}

\end{picture}

}

\def\figinv{

\def\JPicScale{0.8}
\ifx\JPicScale\undefined\def\JPicScale{1}\fi
\unitlength \JPicScale mm
\begin{picture}(120,70)(0,0)
\linethickness{0.3mm}
\qbezier(20,70)(25.19,54.25)(30,54.25)
\qbezier(30,54.25)(34.81,54.25)(40,70)
\linethickness{0.3mm}
\qbezier(20,10)(25.19,25.75)(30,25.75)
\qbezier(30,25.75)(34.81,25.75)(40,10)
\linethickness{0.3mm}
\qbezier(80,70)(90.5,54.44)(90.5,40)
\qbezier(90.5,40)(90.5,25.56)(80,10)
\linethickness{0.3mm}
\qbezier(120,70)(109.5,54.44)(109.5,40)
\qbezier(109.5,40)(109.5,25.56)(120,10)
\put(30,0){\makebox(0,0)[cc]{(a)}}

\put(100,0){\makebox(0,0)[cc]{(b)}}

\end{picture}

}

\def\figclosed{

\def\JPicScale{0.5}
\ifx\JPicScale\undefined\def\JPicScale{1}\fi
\unitlength \JPicScale mm
\begin{picture}(120,70)(0,0)
\linethickness{0.3mm}
\linethickness{0.3mm}
\linethickness{0.3mm}
\qbezier(80,70)(90.5,54.44)(90.5,40)
\qbezier(90.5,40)(90.5,25.56)(80,10)
\linethickness{0.3mm}
\qbezier(120,70)(109.5,54.44)(109.5,40)
\qbezier(109.5,40)(109.5,25.56)(120,10)

\linethickness{0.1mm}
\put(70,40){\line(1,0){60}}

\put(100,0){\makebox(0,0)[cc]{(b)}}

\end{picture}

}

\def\figclose{

\def\JPicScale{0.5}
\ifx\JPicScale\undefined\def\JPicScale{1}\fi
\unitlength \JPicScale mm
\begin{picture}(70,70)(0,0)
\linethickness{0.3mm}
\qbezier(35,70)(45.38,49)(55,49)
\qbezier(55,49)(64.62,49)(75,70)
\linethickness{0.3mm}
\qbezier(5,10)(15.38,31)(25,31)
\qbezier(25,31)(34.62,31)(45,10)
\linethickness{0.1mm}
\put(10,40){\line(1,0){60}}

\put(40,0){\makebox(0,0)[cc]{(a)}}

\end{picture}

}

\def\figcloseb{

\def\JPicScale{0.5}
\ifx\JPicScale\undefined\def\JPicScale{1}\fi
\unitlength \JPicScale mm
\begin{picture}(70,70)(0,0)
\linethickness{0.3mm}
\qbezier(20,70)(30.38,49)(40,49)
\qbezier(40,49)(49.62,49)(60,70)
\linethickness{0.3mm}
\qbezier(20,10)(30.38,31)(40,31)
\qbezier(40,31)(49.62,31)(60,10)
\linethickness{0.1mm}
\put(10,40){\line(1,0){60}}

\put(40,0){\makebox(0,0)[cc]{(b)}}

\end{picture}

}

\def\figclosec{

\def\JPicScale{0.5}
\ifx\JPicScale\undefined\def\JPicScale{1}\fi
\unitlength \JPicScale mm
\begin{picture}(70,80)(0,0)
\linethickness{0.3mm}
\qbezier(50,70)(55.19,38.5)(60,38.5)
\qbezier(60,38.5)(64.81,38.5)(70,70)
\linethickness{0.3mm}
\qbezier(20,10)(25.19,46.75)(30,46.75)
\qbezier(30,46.75)(34.81,46.75)(40,10)

\linethickness{0.1mm}
\put(10,42){\line(1,0){60}}

\put(40,0){\makebox(0,0)[cc]{(c)}}

\end{picture}

}

\def\figclosedrep{

\def\JPicScale{0.5}
\ifx\JPicScale\undefined\def\JPicScale{1}\fi
\unitlength \JPicScale mm
\begin{picture}(120,70)(0,0)
\linethickness{0.3mm}
\linethickness{0.3mm}
\linethickness{0.3mm}
\qbezier(80,70)(90.5,54.44)(90.5,40)
\qbezier(90.5,40)(90.5,25.56)(80,10)
\linethickness{0.3mm}
\qbezier(120,70)(109.5,54.44)(109.5,40)
\qbezier(109.5,40)(109.5,25.56)(120,10)

\linethickness{0.1mm}
\put(70,40){\line(1,0){160}}

\end{picture}

}

\def\figcloserep{

\def\JPicScale{0.5}
\ifx\JPicScale\undefined\def\JPicScale{1}\fi
\unitlength \JPicScale mm
\begin{picture}(70,70)(0,0)
\linethickness{0.3mm}
\qbezier(35,70)(45.38,49)(55,49)
\qbezier(55,49)(64.62,49)(75,70)
\linethickness{0.3mm}
\qbezier(5,10)(15.38,31)(25,31)
\qbezier(25,31)(34.62,31)(45,10)
\linethickness{0.1mm}
\put(10,40){\line(1,0){160}}

\put(70,40){\makebox(0,0)[cc]{$\times$}}

\end{picture}

}

\def\figclosecrep{

\def\JPicScale{0.5}
\ifx\JPicScale\undefined\def\JPicScale{1}\fi
\unitlength \JPicScale mm
\begin{picture}(70,80)(0,0)
\linethickness{0.3mm}
\qbezier(50,66)(55.19,34.5)(60,34.5)
\qbezier(60,34.5)(64.81,34.5)(70,66)
\linethickness{0.3mm}
\qbezier(20,10)(25.19,46.75)(30,46.75)
\qbezier(30,46.75)(34.81,46.75)(40,10)

\linethickness{0.1mm}
\put(10,40){\line(1,0){60}}

\end{picture}

}

\def\figclosegapa{

\def\JPicScale{0.8}
\ifx\JPicScale\undefined\def\JPicScale{1}\fi
\unitlength \JPicScale mm
\begin{picture}(70,70)(0,0)
\linethickness{0.3mm}
\qbezier(25,70)(35.38,49)(45,49)
\qbezier(45,49)(54.62,49)(65,70)
\linethickness{0.3mm}
\linethickness{0.1mm}
\put(20,50){\line(1,0){60}}

\linethickness{0.3mm}
\linethickness{0.3mm}
\qbezier(20,20)(25.19,56.75)(30,56.75)
\qbezier(30,56.75)(34.81,56.75)(40,20)

\linethickness{0.1mm}
\put(30,50){\line(0,1){11}}

\end{picture}

}

\def\figclosegapb{

\def\JPicScale{0.8}
\ifx\JPicScale\undefined\def\JPicScale{1}\fi
\unitlength \JPicScale mm
\begin{picture}(70,70)(0,0)
\linethickness{0.3mm}
\qbezier(15,77)(25.38,56)(35,56)
\qbezier(35,56)(44.62,56)(55,77)
\linethickness{0.3mm}
\linethickness{0.1mm}
\put(10,50){\line(1,0){60}}

\linethickness{0.3mm}
\linethickness{0.3mm}
\qbezier(20,20)(25.19,56.75)(30,56.75)
\qbezier(30,56.75)(34.81,56.75)(40,20)

\linethickness{0.1mm}
\put(30.4,50){\line(0,1){11}}

\end{picture}

}

\def\figclosedrepag{

\def\JPicScale{0.7}
\ifx\JPicScale\undefined\def\JPicScale{1}\fi
\unitlength \JPicScale mm
\begin{picture}(120,70)(0,0)
\linethickness{0.3mm}
\linethickness{0.3mm}
\linethickness{0.3mm}
\qbezier(80,60)(90.5,44.44)(90.5,30)
\qbezier(90.5,30)(90.5,15.56)(80,0)
\linethickness{0.3mm}
\qbezier(120,60)(109.5,44.44)(109.5,30)
\qbezier(109.5,30)(109.5,15.56)(120,0)

\linethickness{0.1mm}
\put(70,45){\line(1,0){60}}

\end{picture}

}

\def\figcloserepag{

\def\JPicScale{0.7}
\ifx\JPicScale\undefined\def\JPicScale{1}\fi
\unitlength \JPicScale mm
\begin{picture}(70,70)(0,0)
\linethickness{0.3mm}
\qbezier(35,70)(45.38,49)(55,49)
\qbezier(55,49)(64.62,49)(75,70)
\linethickness{0.3mm}
\qbezier(5,10)(15.38,31)(25,31)
\qbezier(25,31)(34.62,31)(45,10)
\linethickness{0.1mm}
\put(10,45){\line(1,0){60}}

\end{picture}

}

\def\figclosecrepag{

\def\JPicScale{0.7 }
\ifx\JPicScale\undefined\def\JPicScale{1}\fi
\unitlength \JPicScale mm
\begin{picture}(70,80)(0,0)
\linethickness{0.3mm}
\qbezier(50,76)(55.19,44.5)(60,44.5)
\qbezier(60,44.5)(64.81,44.5)(70,76)
\linethickness{0.3mm}
\qbezier(20,20)(25.19,56.75)(30,56.75)
\qbezier(30,56.75)(34.81,56.75)(40,20)

\linethickness{0.1mm}
\put(10,45){\line(1,0){60}}

\end{picture}

}

\def\fignewrep{

\def\JPicScale{0.6}
\ifx\JPicScale\undefined\def\JPicScale{1}\fi
\unitlength \JPicScale mm
\begin{picture}(110,80)(0,0)
\linethickness{0.1mm}
\put(10,40){\line(1,0){100}}
\linethickness{0.4mm}
\qbezier(40,80)(45.19,69.56)(50,64.75)
\qbezier(50,64.75)(54.81,59.94)(60,60)
\qbezier(60,60)(65.2,60.02)(68.81,58.81)
\qbezier(68.81,58.81)(72.42,57.61)(75,55)
\qbezier(75,55)(77.56,52.3)(82.38,58.31)
\qbezier(82.38,58.31)(87.19,64.33)(95,80)
\linethickness{0.4mm}
\qbezier(30,80)(35.19,69.55)(40,65.94)
\qbezier(40,65.94)(44.81,62.33)(50,65)
\qbezier(50,65)(55.2,67.64)(59.41,66.44)
\qbezier(59.41,66.44)(63.62,65.23)(67.5,60)
\qbezier(67.5,60)(71.3,54.69)(81.53,59.5)
\qbezier(81.53,59.5)(91.76,64.31)(110,80)
\linethickness{0.2mm}
\qbezier(40,10)(45.19,20.44)(50,25.25)
\qbezier(50,25.25)(54.81,30.06)(60,30)
\qbezier(60,30)(65.17,30.02)(71.19,28.81)
\qbezier(71.19,28.81)(77.2,27.61)(85,25)
\qbezier(85,25)(92.8,22.45)(98.81,16.44)
\qbezier(98.81,16.44)(104.83,10.42)(110,0)
\linethickness{0.2mm}
\qbezier(45,10)(60.63,23.06)(69.66,27.88)
\qbezier(69.66,27.88)(78.68,32.69)(82.5,30)
\qbezier(82.5,30)(86.38,27.47)(90.59,20.25)
\qbezier(90.59,20.25)(94.8,13.03)(100,0)
\end{picture}

}

\def\fignewrepmod{

\def\JPicScale{0.6}
\ifx\JPicScale\undefined\def\JPicScale{1}\fi
\unitlength \JPicScale mm
\begin{picture}(110,80)(0,0)
\linethickness{0.1mm}
\put(20,33){\line(1,0){100}}
\linethickness{0.4mm}
\qbezier(40,70)(45.19,59.56)(50,54.75)
\qbezier(50,54.75)(54.81,49.94)(60,50)
\qbezier(60,50)(65.2,50.02)(68.81,48.81)
\qbezier(68.81,48.81)(72.42,47.61)(75,45)
\qbezier(75,45)(77.56,42.3)(82.38,48.31)
\qbezier(82.38,48.31)(87.19,54.33)(95,70)
\linethickness{0.4mm}
\qbezier(30,85)(35.19,74.55)(40,70.94)
\qbezier(40,70.94)(44.81,67.33)(50,70)
\qbezier(50,70)(55.2,72.64)(59.41,71.44)
\qbezier(59.41,71.44)(63.62,70.23)(67.5,65)
\qbezier(67.5,65)(71.3,59.69)(81.53,64.5)
\qbezier(81.53,64.5)(91.76,69.31)(110,85)
\linethickness{0.2mm}
\qbezier(40,20)(45.19,30.44)(50,35.25)
\qbezier(50,35.25)(54.81,40.06)(60,40)
\qbezier(60,40)(65.17,40.02)(71.19,38.81)
\qbezier(71.19,38.81)(77.2,37.61)(85,35)
\qbezier(85,35)(92.8,32.45)(98.81,26.44)
\qbezier(98.81,26.44)(104.83,20.42)(110,10)
\linethickness{0.2mm}
\qbezier(45,10)(60.63,23.06)(69.66,27.88)
\qbezier(69.66,27.88)(78.68,32.69)(82.5,30)
\qbezier(82.5,30)(86.38,27.47)(90.59,20.25)
\qbezier(90.59,20.25)(94.8,13.03)(100,0)
\end{picture}

}

\def\fignogap{

\def\JPicScale{0.6}
\ifx\JPicScale\undefined\def\JPicScale{1}\fi
\unitlength \JPicScale mm
\begin{picture}(110,80)(0,0)
\linethickness{0.1mm}
\put(20,33){\line(1,0){100}}
\linethickness{0.4mm}
\qbezier(40,64)(45.19,53.56)(50,48.75)
\qbezier(50,48.75)(54.81,43.94)(60,44)
\qbezier(60,44)(65.2,44.02)(68.81,42.81)
\qbezier(68.81,42.81)(72.42,41.61)(75,39)
\qbezier(75,39)(77.56,35.8)(82.38,42.31) %
\qbezier(82.38,42.31)(87.19,48.33)(95,64)
\linethickness{0.4mm}
\qbezier(30,85)(35.19,74.55)(40,70.94)
\qbezier(40,70.94)(44.81,67.33)(50,70)
\qbezier(50,70)(55.2,72.64)(59.41,71.44)
\qbezier(59.41,71.44)(63.62,70.23)(67.5,65)
\qbezier(67.5,65)(71.3,59.69)(81.53,64.5)
\qbezier(81.53,64.5)(91.76,69.31)(110,85)
\linethickness{0.2mm}
\qbezier(40,20)(45.19,30.44)(50,35.25)
\qbezier(50,35.25)(54.81,40.06)(60,40)
\qbezier(60,40)(65.17,40.02)(71.19,38.81)
\qbezier(71.19,38.81)(77.2,37.61)(85,35)
\qbezier(85,35)(92.8,32.45)(98.81,26.44)
\qbezier(98.81,26.44)(104.83,20.42)(110,10)
\linethickness{0.2mm}
\qbezier(45,10)(60.63,23.06)(69.66,27.88)
\qbezier(69.66,27.88)(78.68,32.69)(82.5,30)
\qbezier(82.5,30)(86.38,27.47)(90.59,20.25)
\qbezier(90.59,20.25)(94.8,13.03)(100,0)
\end{picture}

}

\def\figpoles{

\def\JPicScale{0.6}
\ifx\JPicScale\undefined\def\JPicScale{1}\fi
\unitlength \JPicScale mm
\begin{picture}(120,80)(0,0)
\linethickness{.6mm}
\put(70,0){\line(0,1){80}}
\linethickness{0.1mm}
\put(20,40){\line(1,0){100}}
\put(90,50){\makebox(0,0)[cc]{$\times$}}

\put(50,50){\makebox(0,0)[cc]{$\times$}}

\put(100,40){\makebox(0,0)[cc]{$\times$}}

\put(40,40){\makebox(0,0)[cc]{$\times$}}

\put(30,20){\makebox(0,0)[cc]{$\times$}}

\put(110,20){\makebox(0,0)[cc]{$\times$}}

\put(70,50){\makebox(0,0)[cc]{$\Uparrow$}}

\end{picture}

}

\def\figdeform{

\def\JPicScale{0.8}
\ifx\JPicScale\undefined\def\JPicScale{1}\fi
\unitlength \JPicScale mm
\begin{picture}(115,80)(0,0)
\linethickness{0.1mm}
\put(20,40){\line(1,0){90}}
\linethickness{0.1mm}
\put(60,0){\line(0,1){80}}
\linethickness{0.6mm}
\put(60,40){\line(0,1){30}}
\linethickness{0.6mm}
\put(60,70){\line(1,0){40}}
\put(65,85){\makebox(0,0)[cc]{$\Imm(p_a^0)$}}

\put(120,40){\makebox(0,0)[cc]{$\Imm(p_a^1)$}}

\put(60,55){\makebox(0,0)[cc]{$\Uparrow$}}

\put(75,70){\makebox(0,0)[cc]{$\Rightarrow$}}

\end{picture}

}

\begin{document}

\vskip 12pt

\baselineskip 24pt

\begin{center}


{\Large \bf Analyticity and Crossing Symmetry of  Superstring Loop Amplitudes}

\end{center}

\vskip .6cm
\medskip

\vspace*{4.0ex}

\baselineskip=18pt

\begin{center}

{\large 
\rm Corinne de Lacroix$^a$, Harold Erbin$^b$, 
Ashoke Sen$^{c}$}

\end{center}

\vspace*{4.0ex}

\centerline{\it $^a$Laboratoire de Physique Th\'eorique, 
Ecole Normale Sup\'erieure,}
\centerline{\it 24 rue Lhomond, 75005 Paris, France}
\centerline{\it  $^b$ASC,
Ludwig--Maximilians--Universit\"at München,}
\centerline{\it Theresienstra\ss{}e 37, 80333 M\"unchen, Germany}
\centerline{ \it  $^c$Harish-Chandra Research Institute, HBNI,}
\centerline{\it Chhatnag Road, Jhusi,
Allahabad 211019, India}

\vspace*{1.0ex}
\centerline{\small E-mail:  lacroix@lpt.ens.fr, harold.erbin@physik.lmu.de,
sen@hri.res.in}

\vspace*{5.0ex}

\baselineskip=18pt

\centerline{\bf Abstract} \bigskip

Bros, Epstein and Glaser proved crossing symmetry of the S-matrix of a theory without massless fields
by using certain analyticity properties of the off-shell momentum space
Green's function in the complex momentum plane. The latter
properties follow from representing the momentum space Green's function as Fourier transform of the position space 
Green's function, satisfying certain properties implied by the underlying local quantum field theory. We prove
the same analyticity properties of the momentum space Green's functions in superstring field theory by directly
working with the momentum space Feynman rules even though the corresponding properties of the position
space Green's function are not known.  Our result is valid to all orders in perturbation theory, but requires, as
usual, explicitly subtracting / regulating the non-analyticities associated with massless particles.
These results can also be used to prove other general analyticity properties of the S-matrix
of superstring theory.

\vfill \eject

\baselineskip=18pt

\tableofcontents

\sectiono{Introduction and summary}

More than 50 years ago, Bros, Epstein and Glaser\cite{BEG1,BEG2,BEG3}  (BEG) 
gave a proof of crossing symmetry in local 
quantum field theories. The goal of this paper will be to generalize the results to superstring 
theory.

We begin by briefly recalling the main ingredients in the proof in \cite{BEG1,BEG2}. We shall use space-time
metric with mostly plus signature.
\begin{enumerate}
\item The position space Green's functions in a $D$ dimensional
local quantum field theory satisfy certain identities
derived from the fact that commutator of local operators vanish outside the light-cone. Using this 
one can show\cite{Bogol,Steinmann,Ruelle,Araki,Araki1} 
that the momentum space {\it amputated} Green's function $G(p_1,\cdots p_n)$ of $n$ external
states, regarded as a function of $(n-1)D$ complex variables after taking into account momentum
conservation $\sum_a p_a=0$, has certain analyticity 
properties.\footnote{Hereafter momentum space Green's
functions will always refer to amputated Green's functions. We do not put any constraint on the spins of the
external particles, but do not explicitly display the Lorentz indices carried by the Green's function.}
If we denote by $P_{(\alpha)}$
the sum over any subset $A_\alpha$ of the $p_a$'s, then the  
Green's function is an analytic function
of the $p_a$'s as long as, 
\be \label{ecrth}
\{\Imm(P_{(\alpha)})\ne 0,   \quad (\Imm(P_{(\alpha)}))^2 \le 0\}, \quad \hbox{or} \quad
\{\Imm(P_{(\alpha)})=0, \, -P_{(\alpha)}^2 < M_\alpha^2\}, \quad \forall A_\alpha\, ,
\ee
where $M_\alpha$ is the  threshold of production 
of any (multi-particle) state in the channel containing the  particles in the set $A_\alpha$.
\item It is easy to see that for massive external particles
the above domain of analyticity does not have any 
overlap with the subspace
of complex momentum space in which the external states are on-shell. Indeed if we write
$p_a=p_{aR}+i p_{aI}$  then the mass shell condition
$p_a^2+m_a^2=0$ implies that $p_{aI}.p_{aR}=0$ and $p_{aR}^2 - p_{aI}^2+m_a^2=0$.
If $p_{aI}$ is non-zero and lie in the forward or backward 
light-cone then $p_{aI}^2\le 0$ and hence $p_{aR}^2<0$  due to the second condition. 
Therefore $p_{aR}$ also lies in the forward or backward light-cone.
However the inner product of  two vectors, each of which
is in the forward or backward light-cone, cannot vanish unless both of them are null, but the $p_{aR}^2<0$
condition shows that $p_{aR}$ is not null. Therefore we cannot satisfy the 
$p_{aR}.p_{aI}=0$ condition. The only possibility is that $p_{aI}=0$ for each $a$, i.e.\ all external
momenta are real. However in this case we cannot satisfy the condition that $P_{(\alpha)}$ is below
the threshold since given any pair of incoming particles (or a pair of outgoing particles) the total
momentum carried by the pair is always sufficient to produce the same pair of particles.
\item Due to this observation the analyticity of the off-shell Green's function in the domain described
above is not by itself sufficient to prove crossing symmetry, since the latter 
involves analytic continuation of {\it on-shell}
four point  Green's function from the physical region of s-channel scattering ($s> 0$, $t,u<0$) to the physical 
region of t-channel scattering ($t> 0$, $s,u<0$) along some path in complex momentum space. 
Nevertheless BEG were able to
show, by using the fact that the shape of the 
domain of analyticity of a function of many complex variables has a restricted 
form\cite{Bros2},
that the actual
domain of holomorphy of $G(p_1,\cdots p_n)$ is bigger than \refb{ecrth}, 
and includes a path that interpolates between physical s-channel
region and physical t-channel region in the momentum space keeping all the external particles
on-shell. 
\item For the proof of crossing symmetry, BEG 
needed to use the analyticity property mentioned in point 1 above only
in a subspace of the complex momentum space in which imaginary components of all 
the external momenta
lie in a two dimensional Lorentzian plane. Without loss of generality we can take this to be the
$p^0$-$p^1$  plane.
\end{enumerate}

Superstring field theory is a quantum field theory with infinite number of fields and non-local
interactions that is designed to reproduce perturbative amplitudes of superstring theory.
A detailed   review of (compactified) 
heterotic and type II string field theories  
can be found in \cite{1703.06410}, but we shall need only minimal information
that will be reviewed
in \S\ref{s1}. We can compute off-shell momentum space Green's functions by summing
over Feynman diagrams, but the non-local nature of the vertices, reviewed in \S\ref{s1}, prevents
us from defining position space Green's functions. Therefore the analogue of operator commutativity at spacelike separations is not obvious and we cannot invoke locality to prove
the analyticity of the momentum space  Green's functions in the domain \refb{ecrth}, as in point 1) above. Instead in this
paper we analyze the analyticity properties by directly examining the singularities of the
off-shell Green's function represented as sum over Feynman diagrams. Our proof of analyticity does not 
extend to the full domain \refb{ecrth}, but to a subspace of this domain where the imaginary
part of the external momenta are restricted to lie in a two dimensional Lorentzian 
plane:\footnote{It may be possible to extend this to a larger domain by applying general
results on functions of many complex variables. However so far by direct analysis of the Feynman
diagrams, we have been able to prove analyticity in the restricted domain \refb{ecr1newint}.}
\ben\label{ecr1newint}
&&\hskip -.58in  \Imm p_a^i=0 \quad \forall \quad a=1,\cdots n, \quad i\ne 0,1, \non\\ &&
\hskip -.6in \{\Imm(P_{(\alpha)})\ne 0,   \quad (\Imm(P_{(\alpha)}))^2 \le 0\},
\quad \hbox{or} \quad
\{\Imm(P_{(\alpha)})=0, \, -P_{(\alpha)}^2 < M_\alpha^2\}, \quad \forall A_\alpha\, .
\een
Furthermore one can show that the analyticity of the  Green's function also extends to all other
points that can be obtained from the ones in \refb{ecr1newint} by Lorentz transformation with
complex parameters.
As mentioned above, the property \refb{ecr1newint} 
of the off-shell Green's function is sufficient to prove crossing
symmetry of the S-matrix using the same argument as used by BEG. The steps leading from
\refb{ecr1newint} to the proof of crossing symmetry only relies on the general properties of functions
of several complex variables and not on the details of the theory that produces the Green's 
functions.\footnote{In fact, our proof applies to any ultraviolet complete theory 
in which the singularities of the integrand in the momentum space Feynman diagrams arise from the
usual poles of the propagator. Therefore our analysis also provides an alternative derivation of the
analyticity properties of the momentum space Green's function in the domain \refb{ecr1newint} for
ordinary renormalizable quantum field theories.}

The analysis of BEG however had one underlying assumption -- that the theory does not have any
massless particles so that the domain of analyticity includes the origin in the space of complex
momenta and is in fact a star shaped region around the origin. When there are massless particles
then there are multi-particle states of arbitrary low energies and therefore the threshold $M_\alpha$
appearing in \refb{ecr1newint} can extend all the way to the origin. A related issue is that in the
presence of massless particles the on-shell
 Green's function has infrared singularities even though in high enough dimensions these 
singularities may not lead to divergences. 
Since string theory has massless states, it also suffers from this problem. We propose two
different ways of addressing this issue. The first is to explicitly remove from the Green's function
contributions where any of the internal propagators is that of a massless particle. This can be done
maintaining ultraviolet finiteness and the resulting contribution can be shown to satisfy
\refb{ecr1newint}. In this case crossing symmetry holds for only this part of the S-matrix element.
The other approach will be to regulate the infrared divergence by adding explicit mass terms for
the massless fields in the {\it gauge fixed action}. This also leads to ultraviolet finite Green's function
satisfying \refb{ecr1newint}. While neither of these approaches show the crossing symmetry of the
full amplitude of superstring theory, what they establish is that the possible
lack of crossing symmetry of
the amplitudes of superstring theory is entirely due to the presence of massless fields -- an effect
that is also present in a local quantum field theory with massless fields. Therefore 
our result \refb{ecr1newint} shows that the inherent non-locality
of string theory encoded in the interaction vertices has no effect on the crossing symmetry of
the amplitude, and string theory behaves like a standard local quantum field theory on this aspect.

Before concluding this section, we would like to mention that even though we have emphasized
the proof of crossing symmetry as the main application of our result, the general result \refb{ecr1newint} 
can be used to prove many other useful results about the analytic structure of on-shell amplitudes in
superstring theory.
In particular if we consider a configuration of external momenta where one particular combination of 
external momenta is allowed to have complex imaginary part keeping all other linearly independent 
combinations real,\footnote{In this case we can always make a Lorentz transformation to make the
imaginary part lie in a given two dimensional Lorentzian plane. Therefore the difference between the
regions \refb{ecrth} and \refb{ecr1newint} becomes irrelevant.}
then once we establish analyticity in the region \refb{ecr1newint}, the domain
of analyticity can be extended to a much larger domain known as the 
Jost-Lehmann-Dyson analyticity 
domain\cite{jost,dyson}. A general proof of this result that relies only on general
properties of functions of many complex variables can be found in \cite{Bros2}.
Using this one can prove various analyticity properties of on-shell amplitudes, {\it e.g.} the
analyticity of the elastic forward scattering amplitude ($t=0$) in the full complex $s$-plane except for
the usual threshold singularities on the real axis (see {\it e.g.} \cite{itzykson}).
The same general result can also be used to determine the domain of analyticity
in the complex $t$-plane for fixed positive $s$\cite{itzykson}.

Some recent discussion on analyticity of the Green's function in D-dimensional theories can be found
in \cite{1608.06402}. Ref.\cite{1302.4290} considered deformations in which a spatial 
component of the external
momenta becomes complex keeping the time components real. However since this leads to space-like
imaginary part of external momenta, the region considered in \cite{1302.4290} does not have any overlap
with the region \refb{ecrth}.

\sectiono{General structure of superstring field theory} \label{s1}

Closed superstring field theory, 
after Lorentz covariant gauge fixing, has infinite number of fields. We shall
label by $\{\phi^\alpha(k)\}$ the momentum space representation of these fields. In a background with
$D$ non-compact space-time dimensions, the action has the
general form\cite{1703.06410}
\ben
S &= & \int {d^D k\over (2\pi)^D}  K_{\alpha\beta}(k) \phi^\alpha(k) \phi^\beta(-k) \non\\  
\hskip -.6in && + \sum_n \int {d^D k_1\over (2\pi)^D} \cdots {d^D k_n\over (2\pi)^D} \, 
(2\pi)^D \delta^{(D)}(k_1+\cdots + k_n) \, 
V^{(n)}_{\alpha_1\cdots \alpha_n} (k_1,\cdots k_n) \\
\non
&& \hspace{8cm} \times \; \phi^{\alpha_1}(k_1)\cdots
\phi^{\alpha_n}(k_n)\, ,
\een
where $K_{\alpha\beta}(k)$ is the kinetic operator that is typically quadratic function of momenta and
$V^{(n)}_{\alpha_1\cdots \alpha_n} (k_1,\cdots, k_n)$ is the off-shell vertex that has the property that
whenever any subset of momenta $k_i$ approach infinity, the dominant factor in the vertex takes the
form $\exp(-c_{ij} k_i.k_j)$ for some matrix $c_{ij}$ with large positive eigenvalues.\footnote{This exponential behaviour is responsible for the non-local behaviour of the vertices and the impossibility of using the position space representation. Since the latter was used in~\cite{Steinmann,Ruelle,Araki,Araki1} to prove the analyticity of the momentum space Green's functions, one needs to find another approach.}
For this reason in computing off-shell Green's functions from this action using Feynman
diagrams,  we always take the integration contours of loop energies to 
run from $-i\infty$ to $i\infty$, although in the interior of the complex plane the contours may be deformed
away from the imaginary axis to avoid poles of the propagators. Similarly
the ends of the integration contours 
of the spatial components of loop momenta will be taken to approach $\pm \infty$ along the real axis. 
This ensures that 
$c_{ij}k_i.k_j$ becomes large and positive as the contour approaches infinity and the loop momentum integrals
are convergent. More generally one can take the loop energy integrals to approach infinity
by remaining within a 45$^\circ$ cone around the imaginary axis and the spatial components of loop
momenta  to approach infinity
by remaining within a 45$^\circ$ cone around the real axis. 
Besides this
$V^{(n)}_{\alpha_1\cdots \alpha_n}(k_1,\cdots , k_n)$ carries a factor of 
$\exp(-c \sum_{i=1}^n m_{\alpha_i}^2)$
for some large positive constant $c$ where $m_\alpha$ is the mass of the field $\phi_\alpha$. 
This ensures convergence in the sum over states in the internal propagators
even though there are infinite number of states.
The vertices are also free from any singularity at finite
points in the complex $\{k_i\}$ planes. Therefore all possible singularities of the  
Green's functions will arise from 
the poles of the propagators. These are simple poles at $k^2+m_\alpha^2=0$.\footnote{In the natural formulation
of superstring field theory action, $K_{\alpha\beta}$ will be the tree level kinetic term and $m_\alpha$'s will be the
tree level masses. However by adding a suitable finite counterterm to $K_{\alpha\beta}$ and subtracting it from
$V^{(2)}_{\alpha\beta}$, we can take the $m_\alpha$'s to be the quantum corrected physical 
masses\cite{1604.01783}. The additional term in $V^{(2)}$ does not carry the exponential suppression factors,
but the exponential suppression factors in the neighboring vertices of the Feynman diagram continue to guarantee
ultra-violet finiteness of individual diagrams.}

Superstring theory has massless states. For reasons explained in the introduction, the kind of questions we would
like to address in this paper requires working with massive theories since in the presence of massless states the on-shell
S-matrix always suffers from infrared singularities and is never fully analytic. 
Therefore the best we can hope for
in string theory is to prove the required analyticity of the  Green's 
functions after separating out the contribution from
the massless states. We can explore two possible ways:
\begin{enumerate}
\item
We can introduce a projection operator $P$ which, acting on the 
space of string fields, projects onto the massless fields. We insert the identity
\be 
(1-P)+P\, ,
\ee
in each propagator and write the contribution from a Feynman diagram as a sum of many terms, where in a given term each
internal propagator
carries either the factor of $P$ or $(1-P)$. Each of these terms is ultraviolet finite due to the exponential suppression
from the vertex. The particular term where all internal propagators carry a factor $(1-P)$ is also infrared finite, and it is for this 
contribution to the Green's
function that we can prove the desired analyticity properties.\footnote{Since the projection operator $P$ 
commutes with the BRST charge, one can show, using a generalization of the analysis described in
\cite{1703.06410,1609.00459}, that the generating functional of these subtracted Green's functions satisfies 
appropriate 
BV master
equations capturing the full gauge invariance of the theory.
When all the external states are massless, these
contributions to the
Green's functions are precisely the vertices of the Wilsonian 
effective field theory of massless fields obtained by integrating
out the massive fields\cite{1703.06410,1609.00459}, but the definition given
here also applies for massive external states. \label{fo2}} 
The rest of the terms have one or more massless internal propagators
and may
suffer from infrared singularities. However these are the usual infrared singularities associated with massless states
and would occur also in local quantum field theories. Therefore establishing the desired analyticity of the infrared 
safe part would show that  there is no distinction between
the analytic properties of the amplitudes in string theory and that of a local quantum field theory with massless
states.

As mentioned in footnote \ref{fo2}, the off-shell amplitude with infrared subtractions satisfies appropriate
form of BV master equation that encodes gauge invariance of the full amplitude. Nevertheless the subtracted
amplitudes by themselves are not gauge invariant.
Given this, one could
wonder whether our result has any gauge invariant content. For definiteness let us consider a 
specific class of subtraction procedure. It is known that for a given string compactification one can
construct an infinite family of superstring field theories by making different choices of local
coordinate systems on the world-sheet. It is also known that these different superstring field theories
are related to each other by field redefinition. We can now develop perturbation theory in each of
these formulations of superstring field theory by imposing Siegel gauge condition, but since 
Siegel gauge in one formulation does not correspond to Siegel gauge in another formulation, the
Green's functions computed in different formulations of the theory will correspond to different gauge
choices. If we now subtract
the infrared divergent part following the procedure suggested above
the results will be different in different formulations. 
However the statement that is independent of the gauge choice is that in each of these
formulations the Green's functions so defined will be analytic in the domain \refb{ecr1newint}. 

\item 
Alternatively one could add to the action explicit mass terms for the massless fields in the gauge fixed action. This
will break BRST invariance and decoupling of unphysical states in the S-matrix, but will generate off-shell Green's
functions free from infrared singularities. Our general analysis of analyticity properties given below will hold for these
regularized amplitudes as well. The infrared singularities will reappear in the massless limit.

In higher dimensions one can have better control on the situation as follows. 
If we regulate the theory by adding explicit mass term of order $m$ to the massless
particles, then the dependence of the amplitude on $m$ will be via positive powers of $m$, possibly
multiplied by non-analytic terms like $\ln m$. 
Therefore the regulated amplitude, that has the analyticity and crossing
properties, differs from the actual amplitude by a small amount if we take $m$ to be small.
\end{enumerate}

In the following we shall assume that all the states propagating in the internal propagators have non-zero
mass.
One may be surprised at the ability to modify string theory amplitudes by subtracting / regularizing infrared singular
parts and still preserve ultraviolet finiteness. This follows clearly from the exponential suppression factors in the
vertices for large {\it imaginary} energy and real spatial momenta. 
What these subtracted / regularized amplitudes lack however is good behavior at large {\it
real} energy.
Individual
Feynman diagrams diverge rapidly for large real energies of the external states due to the  exponential factors
in the vertices. Only after adding the contributions from different diagrams we get sensible high energy behavior due
to delicate cancellation between different terms. 
Individual pieces / regularized amplitudes will lack this cancellation.

The analysis of \cite{BEG1,BEG2} also requires that the external particles are stable. We shall take the external 
particles to be either massless states, or BPS states or stable non-BPS states of superstring theory.

\sectiono{Analyticity property of the off-shell Green's functions} \label{s2}

\begin{figure}
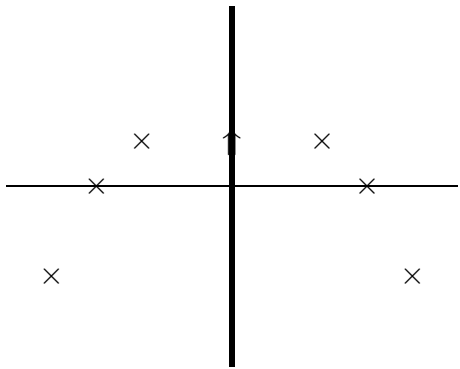


\begin{center}

\figpoles 

\end{center}

\caption{Distribution of poles in a complex loop energy plane for vanishing external
momenta.
The poles are situated symmetrically about the imaginary axis. The vertical positions
of the poles are determined by the values of purely imaginary loop energies along other loops, and
the horizontal positions are determined by the spatial components of all loop momenta and
masses which are all real. 
The loop energy integration contour is taken to be from $-i\infty$
to $i\infty$ along the imaginary axis as shown by the thick vertical line. As we deform the
external momenta away from zero the pole positions move, and if any pole approaches the
integration contour  we have to deform the integration contour to avoid the poles.
\label{figpole}}

\end{figure}

Let $p_a$ be the momentum of the $a$-th external particle for $a=1,\cdots n$, counted as positive if ingoing and
negative if outgoing. In $D$ space-time dimensions the complex momenta 
$(p_1,\cdots p_n)$ satisfying $\sum_a p_a=0$ span a $(n-1)D$ dimensional complex manifold
$\CCC^{(n-1)D}$ and the Green's function $G(p_1,\cdots p_n)$ is a function of these $(n-1)D$ complex
variables, obtained by summing over Feynman diagrams of superstring field theory. 
We adopt the following procedure for computing the loop momentum integrals in a Feynman diagram.
At the origin where all the external momenta vanish ($p_a=0$), we take each of the loop energy integrals to run
from $-i\infty$ to $i\infty$ along the imaginary axis and each spatial component of loop momenta to run from
$-\infty$ to $\infty$ along the real axis. In this case each internal line carries imaginary energy and real spatial
momenta, making the denominator factor $(\ell^2+m^2)$ of the propagator carrying momentum $\ell$ 
strictly positive. Therefore the integrand does not have any singularity. Furthermore due
to exponential suppression from the vertices the integrals are convergent as loop energies approach 
$\pm i\infty$ and spatial components of loop momenta approach $\pm\infty$. Therefore the Green's
function is non-singular at the origin. Fig.~\ref{figpole} shows the distribution of poles at 
$\ell^0=\pm\sqrt{\vec\ell^2+m^2}$ in a complex
loop energy plane for vanishing external momenta.

As the $p_a$'s move away from the origin of $\CCC^{(n-1)D}$, 
we can continue to use the same loop energy and
loop momentum integration contours as long as the poles are away from the integration contours. When any of
the poles approaches the integration contour we need to deform the contour away from the poles keeping
its ends fixed at $\pm i\infty$ for loop energies and $\pm \infty$ for loop momenta. When a pair of poles
approach an integration contour from opposite sides so that we can no longer deform the contour away
from the poles, the integral itself becomes singular\cite{sterman}. 
Since initially the poles are located at finite distance away from the integration contour, 
the singularities of the Green's function will be located at finite distance away from the origin.
Therefore there exists a connected region in $\CCC^{(n-1)D}$ containing
the origin in which the Green's function is a complex analytic function of the external momenta.
\cite{1604.01783} explored a subspace of $\CCC^{(n-1)D}$ in which we keep all the spatial components of the external
momenta real and take all the external energies to be given by a single complex number $\lambda$ 
multiplying real
numbers. Green's functions were shown to be
analytic as long as $\lambda$ remains in the first quadrant, and
the physical scattering amplitudes were defined as the $\lambda\to 1$ limit of these Green's functions from
the first quadrant.
Our goal will be to explore a larger domain in $\CCC^{(n-1)D}$ in which the
Green's function remains a complex analytic function of external momenta. 

One property of the domain that
follows automatically is its invariance under complex Lorentz transformation -- Lorentz transformation
with complex parameters. Indeed, when we make 
a particular complex Lorentz transformation of the external momenta, we can apply identical transformation
on the loop momentum integration contours. This does not change any of the denominators of the propagators,
or any of the exponential factors needed for ensuring convergence of the integral at infinity. Therefore if the
Green's function was analytic for the original values of the external  momenta, it will remain analytic for the new
values of the external momenta.

\begin{figure}
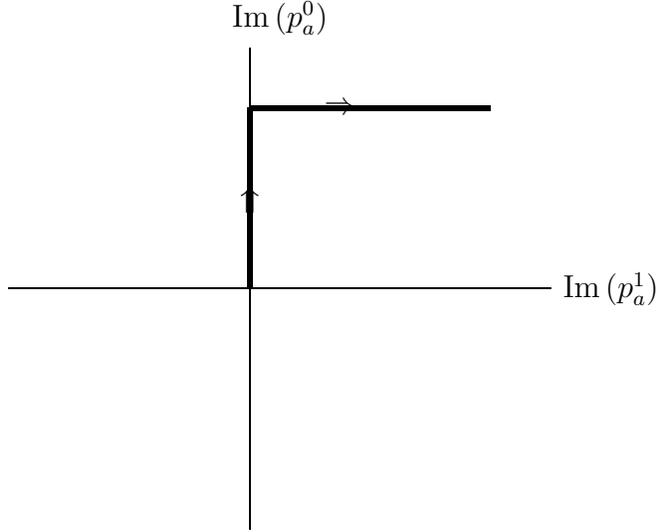


\begin{center}

\figdeform 

\end{center}

\caption{A schematic representation of the deformation of the external momenta that we follow in the
$\Imm(p_a^1)$-$\Imm(p_a^0)$ plane. In the first step we deform all components of momenta except
$\Imm(p_a^1)$ to their desired values along a straight line. This has been represented by the thick 
vertical line. In the second step we deform $\Imm(p_a^1)$ to their desired values along a straight line
keeping fixed all other components. This has been shown by the thick horizontal line.
\label{figdeform}}

\end{figure}

Our strategy for establishing analyticity of the Green's function for a given set of values of $\{p_a\}$ will be to
find a path connecting the  origin to $\{p_a\}$ and show that as we \cdnote{move} the external momenta 
along that path, we can continuously deform the loop momentum
integration contours  avoiding the
poles of the propagators. This will establish the analyticity of the Green's function at $\{p_a\}$. We shall 
carry out this deformation in two steps. First we deform all real components of all the external 
momenta and the
$\Imm(p_a^0)$'s to the desired values along a straight line, keeping 
$\Imm(p_a^1)$'s fixed at 0
for each $a$. In the second step we deform the $\Imm(p_a^1)$'s to their desired values along a straight line
keeping all other components fixed. This has been shown schematically in 
Fig.~\ref{figdeform}.

\subsection{Complex energy but real spatial momenta} \label{s2.1}

Let us define
\be \label{ecr0}
P_{(\alpha)} \equiv \sum_{a\in A_\alpha} p_a\, ,
\ee 
for any proper subset $A_\alpha$ of $\{1,\cdots n\}$ other than the empty set. 
We shall now show, generalizing the arguments in \cite{1604.01783}, that the  
off-shell Green's function is free from singularity if
\ben \label{ecr1}
&& \Imm p_a^i=0 \quad \forall \quad a=1,\cdots n, \quad 1\le i\le (D-1), \non\\
&& \left\{\Imm(P_{(\alpha)}^0)\ne 0 \quad \hbox{or} \quad
\Imm(P_{(\alpha)}^0)= 0, \, -P_{(\alpha)}^2 < M_\alpha^2\right\} 
\quad \forall A_\alpha\subset \{1,\cdots n\}\, ,
\een
where $M_\alpha$ is the threshold invariant mass  
for producing any set of intermediate states in the
collision of particles carrying total momentum $P_{(\alpha)}$.
We shall denote by $S$ the set of points in $\CCC^{(n-1)D}$ satisfying 
\refb{ecr1}.

Given any point $(p_1,\cdots p_n)$ in the complex momentum space we define the Green's function 
$G(p_1,\cdots p_n)$ as sum over Feynman diagrams, but we need to specify the contours along which the loop
momenta are integrated. In the analysis of this subsection, the 
spatial components of the loop momenta will always be integrated along the real axis.
The integration contours for the loop energies are chosen as follows.
We draw a straight line connecting the origin to $(p_1,\cdots p_n)$ in $\CCC^{(n-1)D}$ satisfying 
\refb{ecr1}, parametrized by
a real parameter $\lambda$. As we deform $\lambda$ away from zero, we continue to integrate the independent loop
energies along the imaginary axes as long as the locations of the poles of the propagators do not approach the
contours. However if any of the poles of the propagator
approach the integration contour, then we deform the loop energy integration 
contours away so as to avoid the singularity, keeping their ends
tied at $\pm i\infty$ in order to ensure convergence of the integral at infinity. When a contour
is pinched by two singularities approaching from opposite sides so that it is no longer possible to
avoid the singularities, we may run into a singularity of the integral.\footnote{Even this may not be a genuine singularity
since we may still be able to deform the spatial components of the loop momenta into the complex plane, but we shall not
explore this possibility.} 
Our goal will be to show that this does not happen.

\begin{figure}
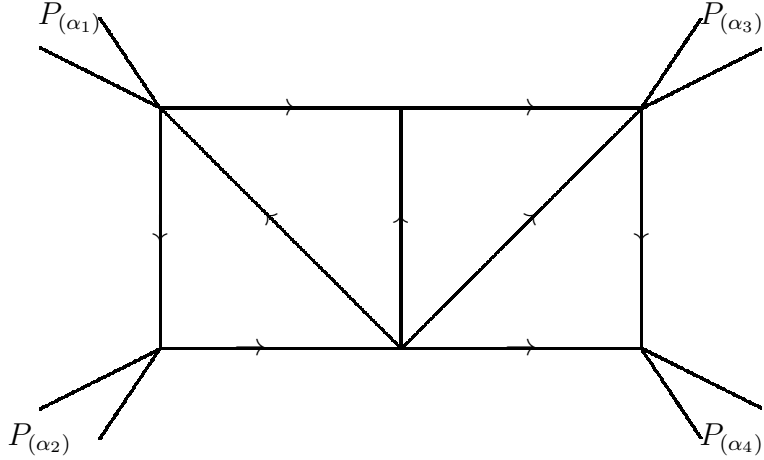


\begin{center}

\figreduced 

\end{center}

\caption{Example of a reduced diagram. The arrows denote the direction of flow of energy along the
on-shell internal propagators. This reduced diagram does not correspond to a pinch singularity since by
following the
arrows we can go around a closed loop in the lower left triangle.\label{figreduced}}

\end{figure}

To prove this we assume the contrary, i.e. that there is a pinch singularity of the loop energy integrals 
during this 
deformation, and show that there is a contradiction. At the pinch certain propagators are 
forced to be on-shell since the corresponding loop energy integration contours cannot be
deformed away from the singularity. 
We now associate to this pinch a reduced diagram by 
contracting to points all propagators that are not forced to be on-shell at the pinch. 
The vertices of the reduced diagram are called reduced vertices. 
The total external momentum entering a given reduced vertex will be given by one of the 
$P_{(\alpha)}$'s.
If $\ell$ is the momentum carried by one of the internal propagators of the reduced diagram and 
$m$ is the mass of the particle propagating in the propagator, 
then the poles of the propagator are located at
\be
\ell^0 = \pm \sqrt{\vec \ell^2 + m^2}\, .
\ee
We shall draw an arrow on the propagator along the direction of $\ell$ if the pole corresponding to
$\ell^0 = \sqrt{\vec \ell^2 + m^2}$ approaches the integration contour, and draw an arrow on the propagator 
opposite to the direction of $\ell$ if the pole corresponding to
$\ell^0 = -\sqrt{\vec \ell^2 + m^2}$ approaches the integration contour. 
An example of a reduced diagram has been shown in 
Fig.~\ref{figreduced}.

We shall now prove that in order that the loop energy integration contour is pinched, we cannot have 
an oriented closed loop in the reduced diagram. Fig.~\ref{figreduced} contains such an oriented closed
loop in the lower left hand triangle. To prove this, let us assume that the reduced diagram has such an
oriented closed loop, and let us label the independent loop momenta such that one loop momentum flows along
the particular oriented closed loop. If we denote by $k$ the loop momentum flowing in this loop
along the arrow, 
by $\{\ell_r\}$, $r\in \AAA$ the momenta along the arrows carried by individual propagators along the loop
and by $\{m_r\}$, $r\in\AAA$ the masses of the particles flowing along these propagators, then:
\begin{enumerate}
\item We have $\ell_r = k + L_r$ for $r\in\AAA$ 
where $L_r$ is a linear combination of the external momenta and other loop
momenta.
\item No propagator outside the set $\AAA$ carries the momentum $k$.
\item Due to the structure of the arrows it follows that the poles which approach the $k^0$ integration contour are
the ones at $\ell_r^0=\sqrt{\vec \ell_r^2+m_r^2}$.
\end{enumerate}
Now at $\{p_a=0\}$, all loop momentum integration contours run along the imaginary axis. As $k^0$ varies from
$-i\infty$ to $i\infty$
along the imaginary axis keeping the other loop momenta fixed, each of the $\ell_r^0$'s 
for $r\in\AAA$ vary from $-i\infty$ to $i\infty$.
Therefore the poles at $\ell_r^0=\sqrt{\vec \ell_r^2+m_r^2}$ lie to the right of the $k^0$ 
integration contour. As we deform the external
momenta away from the origin, we deform the $k^0$ integration contour to avoid the poles of the propagators, but since
the poles do not cross the integration contour they continue to lie on the right of the $k^0$ axis. Therefore when we
approach the supposed pinch singularity where the poles at $\ell_r^0=\sqrt{\vec \ell_r^2+m_r^2}$ approach the
$k^0$ integration contour, they all approach the contour from the right. As a result we can deform the $k^0$ integration
contour away from the poles, showing that this is not a pinch singularity. This proves that in order that the reduced
diagram represents a pinch singularity, we cannot have an oriented closed loop in the diagram.

\begin{figure}
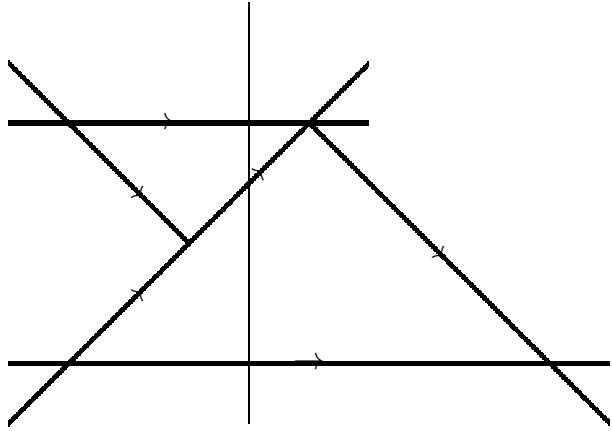


\begin{center}

\figarrow

\end{center}

\caption{A reduced diagram with no oriented closed loop and the arrows drawn so that they always point
to the right. Therefore the energies flow from left to right along all the propagators. The ordering of the
vertices can sometimes be ambiguous, {\it e.g.} any one of the two left-most vertices could be drawn to the
left of the other. \label{figarrow}}

\end{figure}

Absence of oriented closed loop in the diagram allows us to assign a partial ordering of the vertices of a reduced
diagram: if two vertices are connected by a propagator, then the vertex to which the arrow is directed is
drawn to the
right of the vertex from which the arrow originates. Since  $\ell^0=\sqrt{\vec \ell^2+m^2}$ 
is positive if $\ell$ is the momentum carried by the propagator along the arrow, this 
partial ordering implies that  the energy flows from the left to the right. Fig.~\ref{figarrow} shows a particular
reduced diagram drawn in this manner.
If we now draw a vertical line through the propagators 
that divides the diagram into two parts, as shown by the thin vertical line in Fig.~\ref{figarrow},
then across the vertical line there will be on-shell
particles moving from left to the right, carrying total momentum given by one of the $P_{(\alpha)}$'s. If we 
denote by the set $\BBB$ the subset of internal lines intersected by the vertical line, we have
\be
P_{(\alpha)}^0 = \sum_{r\in \BBB} \sqrt{\vec \ell_r^2 + m_r^2}, \quad \vec P_{(\alpha)} = \sum_{r\in \BBB} \vec \ell_r\, .
\ee
Since the spatial components $\vec\ell_r$ are
real along the integration contour, $P_{(\alpha)}$ must be real, and
furthermore since $P_{(\alpha)}$ is the sum of the momenta carried by a set of on-shell particles, we must have
$-P_{(\alpha)}^2 \ge M_\alpha^2$. This contradicts \refb{ecr1}. This shows that
our initial assumption, that there exists a pinch singularity during the deformation of the external momenta from
the origin to a point in the set $S$ defined in \refb{ecr1}, must be wrong. This in turn shows that the Green's function
is analytic at points inside the set $S$ described in \refb{ecr1}.

It should also be clear from the above discussion that the points in the set 
$S$ lie in the interior of the domain of analyticity since 
the integration contours remain at finite distance away from the pinches -- even when $P_{(\alpha)}$ is real
but satisfies the strict inequality $-P_{(\alpha)}^2 < M_\alpha^2$.
Therefore
small deformations in the
external momenta, producing small changes in the momenta carried by the internal propagators,
will not produce a pinch. This shows that for
any point $Q$ in the set $S\subset\CCC^{(n-1)D}$,
we can draw a sufficiently small open neighborhood of
$\CCC^{(n-1)D}$ containing $Q$ where the analyticity property holds. 

\subsection{Complex momenta in two dimensional Lorentzian plane} \label{s2.2}

\begin{figure}
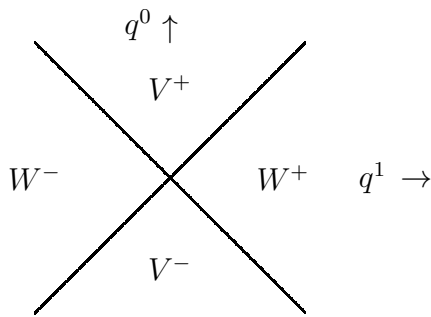


\begin{center}

\figzero

\end{center}

\caption{The four regions $V^\pm$ and $W^\pm$ in the two dimensional momentum plane. \label{f0}}

\end{figure}

We shall now consider a more general subspace of the full complex momentum space in which we allow only the
time component $p^0$ and one spatial component (say $p^1$) of the external momenta to be complex, but take
all the remaining components to be real. We shall use the symbol $p_\parallel$ for $(p^0,p^1)$ and $p_\perp$
for $(p^2,\cdots p^{D-1})$ and label $p$ as $(p_\parallel,p_\perp)$. We also denote by the subscripts $R$ and $I$ the
real and imaginary parts of any quantity. Therefore we have
\be 
p = p_R + i \, p_I = (p_{\parallel R}, p_{\perp R})+ i\, (p_{\parallel I}, 0)\, .
\ee
We shall divide the $(p^0_R, p^1_R)$ and $(p^0_I, p^1_I)$ planes into four quadrants as follows:
\be 
V^+: q^0 > |q^1|, \quad V^-:q^0 < - |q^1|, \quad 
W^+: q^1 > |q^0|, \quad W^-:q^1 < - |q^0|\,,
\ee
where $q$ stands for either $p_R$ or $p_I$.
This has been shown in Fig.~\ref{f0}.

Let us define, as before,
\be \label{ecr0new}
P_{(\alpha)} \equiv \sum_{a\in A_\alpha} p_a\, ,
\ee 
for any proper subset $A_\alpha$ of $\{1,\cdots n\}$ other than the empty set. 
We shall now show that the  
off-shell Green's function is free from singularity if
\ben\label{ecr1new}
&& \Imm p_a^i=0 \quad \forall \quad a=1,\cdots n, \quad i\ge 2, \non\\
&& P_{(\alpha)\parallel I}\in V^\pm\cup (\p V^\pm -0) \quad \hbox{or} \quad 
\{P_{(\alpha)\parallel I}=0, \, -P_{(\alpha)}^2 < M_\alpha^2\},
\quad \forall A_\alpha\subset \{1,\cdots n\}\, ,
\een
where $M_\alpha$ is the threshold invariant mass  
for producing any set of intermediate states in the
collision of particles in the set $A_\alpha$.

We begin at the origin $\{p_a=0\}$ and first deform the external momenta to 
a point where the components $p_a^0$, $p_a^i$ for $i\ge 2$ and $\Rea(p_a^1)$ reach their desired
values keeping $\Imm(p_a^1)=0$. The analysis of the previous subsection tells us that along this
deformation we do not encounter any singularity. In the second step we deform only the imaginary 
parts of $p_a^1$ keeping all other components fixed along a straight line till we reach the desired value.
It is easy to see that all along the path the external momenta will satisfy \refb{ecr1new}.
Our goal will be to show that during this deformation we do not encounter any singularity.

Before we proceed to give the detailed proof, let us outline the main steps in this analysis.
As before we denote by $\{k_s\}$ the loop momenta of the reduced diagram and by $\{\ell_r\}$ the
momenta carried by individual propagators. Our first task will be to understand where the poles are in 
the $k_s^1$-plane and how they could
possibly pinch the $k_s^1$ contour. Then, we show that at the pinch,
it is possible to order the vertices of the reduced
diagram such that $\Imm(\ell_r^1)$ always flows from the left to the right. We then consider a vertical cut through the
reduced diagram, and show that the total momentum flowing from the left to the right  lie in the domain
$W^+$ in Fig.~\ref{f0}, violating the condition that the external momenta lie within the domain
given in \refb{ecr1new}.
This in turn establishes that the $k_s^1$ integration contours cannot be pinched as long as the external
momenta lie within the domain \refb{ecr1new}, establishing the analyticity of the off-shell Green's function
in this domain.

Let $\{k_s\}$ denote the independent loop momenta flowing in a given Feynman diagram. 
At the end of the first deformation we would have reached a non-singular configuration in which
$k_s^1,\cdots k_s^{D-1}$ would have been integrated along the real axis, but for each value
of these spatial components of loop momentum, $k_s^0$ would have been integrated along some
complicated contour. During the second deformation we choose the integration contours as follows.
Given any point on the integration contour at the end of the first deformation, describing the values
of all components of all loop momenta, we allow the point to move to a new point differing only in
the values of $\{\Imm(k_s^1)\}$. This means that we
continue to integrate $k_s^2,\cdots k_s^{D-1}$
along the real axis, and use the same $k_s^0$ contours as at the end of the first deformation for given
values of $k_s^i$ for $i\ge 2$ and $\Rea(k_s^1)$, but allow the $k_s^1$ contours to be deformed along
the imaginary direction. Put another way, if we project the $k_s^1$ contours in each of the loops to the real
axis, we get back the integration contours obtained at the end of the first deformation.
Our goal will be to show that we can avoid singularities of the integrand by appropriate choice of the
$k_s^1$ contours.

\begin{figure}
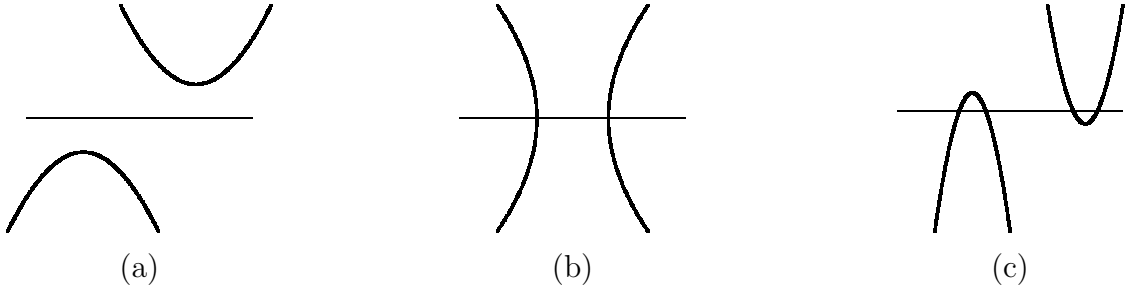


\begin{center}

\hbox{\figclose  \hskip -.4in 
\figclosed \hskip 1in
\figclosec}

\end{center}

\caption{Possible form of the singular loci of the propagator carrying momentum $\ell$ 
in the $\ell_r^1$ plane. The thin horizontal lines denote real $\ell_r^1$ axis. Transitions from the
form (a) or (c) to form (b) could take place by the two branches in (a) or (c) touching each other at $\ell_r^1=0$.
\label{figclose}}

\end{figure}

Before we carry out the deformation we shall try to gain some understanding of how the singularities of the
integrand are distributed in the complex 
$k_s^1$ plane at the end of the first deformation which is the starting point
of the second deformation.  
For this let us focus on a particular loop momentum $k_s$ keeping fixed all other loop momenta.
We denote by $\BBB_s$ the set of internal propagators through which the loop momentum
$k_s$ flows. 
Let $r \in \BBB_s$ index the individual internal propagators appearing in the $s$-th loop, such that $\{\ell_r\}$ and $\{m_r\}$ denote the momenta and masses carried by these propagators along the direction
of flow of $k_s$. Then we have the following
relations between the $\{\ell_r\}$'s and $\{k_s\}$:
\begin{enumerate}
\item We have 
\be \label{ecrdefLr}
\ell_r = k_s + L_r, \quad \hbox{for} \quad r\in \BBB_s\, ,
\ee 
where $L_r$ is a 
linear combination of the external momenta and other loop
momenta. Since at the end of the first deformation
all other loop momenta as well as external momenta have real components along 1-direction,
$L_r^1$ (as well as $L_{r\perp}\equiv (L_r^2,\cdots L_r^{D-1})$) are real.
\item No propagator outside the set $\BBB_s$ carries the momentum $k_s$.
\item For fixed values of other loop momenta, the integration over $k_s^1$ contour 
is along the real axis. Due to reality of $L_r^1$,
this implies that as $k_s^1$ varies from $-\infty$ to $\infty$ along the real axis, all the $\ell_r^1$'s also vary from $-\infty$
to $\infty$ along the real axis.
\end{enumerate}

\begin{figure}
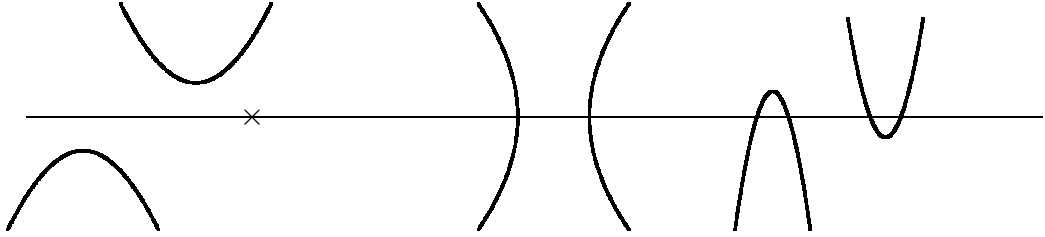


\begin{center}

\hbox{\figcloserep  \hskip -.5in 
\figclosedrep 
\figclosecrep}

\end{center}

\vskip -.3in

\caption{This figure represents, for a given real value of $k_s^1$ denoted by $\times$, the 
singular loci in the complex $k_s^1$ plane from different propagators at the end of the deformation
described in \S\ref{s2.1}.  
The thin horizontal line denotes real $k_s^1$ axis. 
As $k_s^1$ denoted by $\times$ 
moves along the real axis, the shapes of the singular loci change 
so as to make way for the integration
contour to pass all the way from $-\infty$ to $\infty$ without encountering a singular locus.
\label{figcloserep}}

\end{figure}

It will be useful to understand the location of the pole of the propagator carrying momenta $\{\ell_r\}$ in the 
complex $k_s^1$ plane.  The condition for the propagator carrying momentum $\ell_r$ to be on-shell is given by
\be \label{ecr22}
\ell_r^1 = \pm i\, \sqrt{\ell_{r\perp}^2
- (\ell_r^0)^2 + m_r^2}  = \pm i\, \sqrt{(k_{s\perp}+L_{r\perp})^2 - ( k_s^0+L_r^0)^2 + m_r^2}
\quad \hbox{for}\quad r\in \BBB_s
\, .
\ee
As we integrate over the $k_s^0$ contour for fixed values of other loop momentum components,
this generates a
curve describing positions of the poles at $\ell_r^2+m_r^2=0$ 
in the complex $\ell_r^1$ plane via \refb{ecr22}. 
We shall call them the singular loci. 
Since the ends of the $k_s^0$ contours are fixed at $\pm i\infty$,  \refb{ecr22} shows that the 
singular loci in the complex $\ell_s^1$ plane also approach $\pm i\infty$ at the
two ends. However they
could go from $\pm i\infty$ to $\pm i\infty$ or $\pm i\infty$ to $\mp i\infty$. These different situations are shown in 
Fig.~\ref{figclose}(a), (b) and (c).  In all these figures 
the singular loci are related by a reflection around the
origin ($\ell_r^1\to -\ell_r^1$). 
The location of these poles in the $k_s^1$ plane can be found using \refb{ecrdefLr}.
The $k_s^1$ plane will have several such singular loci, one pair for each $\ell_r$,
shifted along the real axis by $L_r$. This has been shown in Fig.~\ref{figcloserep}.

\begin{figure}
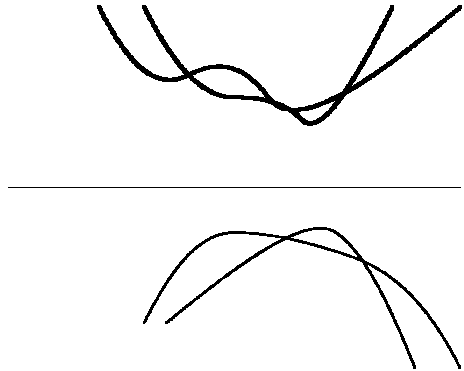


\begin{center}

\fignewrep

\end{center}

\vskip -.2in

\caption{An economical representation of the relevant information of
Fig.~\ref{figcloserep} in which we plot for each real $k_s^1$, only the points
of the singular loci that have the same $\Rea(k_s^1)$. In other words we draw a vertical line through $\times$ in
Fig.~\ref{figcloserep} and plot only the points where it intersects the singular loci. This allows us to capture the
relevant information for all $\Rea(k_s^1)$ in one graph instead of having to plot separate graphs like 
Fig.~\ref{figcloserep}
for different positions of $\times$ on the real axis. Since the singular locus always keeps away from $\times$ in Fig.~\ref{figcloserep}, in this new
representation the curves will never cross the real axis. For reasons to be explained 
in the text, the curves above the real
axis have been drawn thicker than the ones below the real axis.
\label{fignewrep}}

\end{figure}

Recall now that the picture given in Fig.~\ref{figcloserep} is valid for a particular value of $k_s^1$ (denoted by $\times$ in Fig.~\ref{figcloserep}) since the
singular loci depend on the integration contour over $\ell_s^0$ and these in turn can vary as we vary $k_s^1$.
Indeed, otherwise Fig.~\ref{figcloserep} would imply that as we integrate $k_s^1$ along the real axis we hit the
singular loci at the points where the singular loci intersect the real $k_s^1$ axis. This is inconsistent with the 
assertion that at the end of the first deformation 
the integration contours are chosen so as to avoid all singularities of the integrand.
What this means is that as $k_s^1$ approaches one of the points where a singular locus hits 
the real axis, we deform the
$k_s^0$ integration contour so that the singular locus moves away (becoming a configuration like the one in
Fig.~\ref{figclose}(a)), making way for the $k_s^1$ contour to pass through. During this deformation new obstructions
may appear elsewhere on the real $k_s^1$ axis but this does not affect us. The ability to always carry out 
a deformation of the $k_s^0$ contour to make  way for the $k_s^1$ contour to pass through follows from the result of
subsection \S\ref{s2.1} that we do not encounter any singularity during the deformations considered there. This
suggests that the representation of the singular loci as shown in Fig.~\ref{figcloserep} contains too much redundant
information -- for a given value of $k_s^1$ on the real axis, 
the relevant information should contain the locations of the poles in the $k_s^1$ plane whose real part coincides with
the chosen value of $k_s^1$. This has been shown in  
Fig.~\ref{fignewrep}, where we plot, for given
$\Rea(k_s^1)$, the points where the vertical line drawn through $k_s^1$ intersects the 
singular loci in Fig.~\ref{figcloserep}. In this figure the singular loci never cross the real
axis. Note that we have used thicker lines to label the curves above the real axis and
thinner lines to label the curves below the real axis. The points on the 
thicker curves have positive
$\Imm(k_s^1)$ and $\Imm(\ell_r^1)$ and the thinner curves have negative
$\Imm(k_s^1)$ and $\Imm(\ell_r^1)$.

\begin{figure}
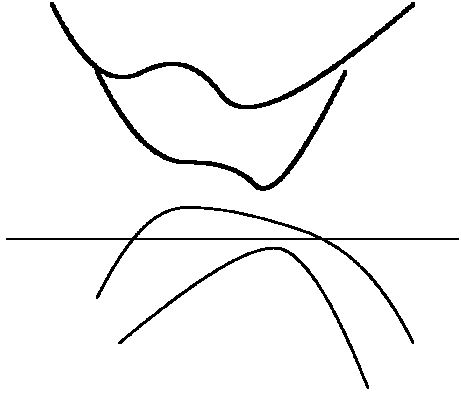


\begin{center}

\fignewrepmod

\end{center}

\caption{Fate of Fig.~\ref{fignewrep} after switching on $\Imm(p_a^1)$. 
Different  singular loci associated with different propagators 
undergo different amounts of vertical shift. In this figure it is no longer possible 
to choose the $k_s^1$ integration contour along the real axis, but it is possible to
find a deformed contour from $-\infty$ to $\infty$ that avoids the singularities.
\label{fignewrepmod}}

\end{figure}

So far we have been considering the end point of the first deformation. Now let us consider the
effect of deforming the imaginary parts of the $p_a^1$'s keeping all other components of
external momenta
fixed. Our goal will be to examine the fate of the singular loci in $k_s^1$ plane for fixed values of
other loop momenta. Now as discussed before, during the second deformation we allow the
integration contours of different loop momenta $\{k_{s'}\}$  
to shift by a change in $\Imm(k_{s'}^1)$, without changing
any other components. The effect of the deformation of $\Imm(p_a^1)$ and  $\Imm k_{s'}^1$
for $s'\ne s$ 
will be to change some of the $L_r$'s appearing in \refb{ecrdefLr} by 
adding 
a constant imaginary part to $L_r^1$. 
This will have the effect of different vertical shifts of different 
singular loci in Fig.~\ref{fignewrep}. As shown in Fig.~\ref{fignewrepmod},
this may now introduce obstruction on the real axis,
preventing us from integrating $k_s^1$ along the real axis. 
As long as there is a vertical gap in the singular loci, we can deform the $k_s^1$ contour
along the imaginary direction and make it pass through the gap without hitting a singularity.
Since we have specified that during this deformation the $k_s^0$ integration contour remains unaltered
for given $\Rea(k_s^1)$, the curves in Fig.~\ref{fignewrepmod} do not
change as we deform the $k_s^1$ contour this way.
If at some point the vertical gap closes, then
we cannot draw the $k_s^1$ contour without hitting a singularity. This has been shown in
Fig.~\ref{fignogap}. At this point we can make the following observations:
\begin{figure}
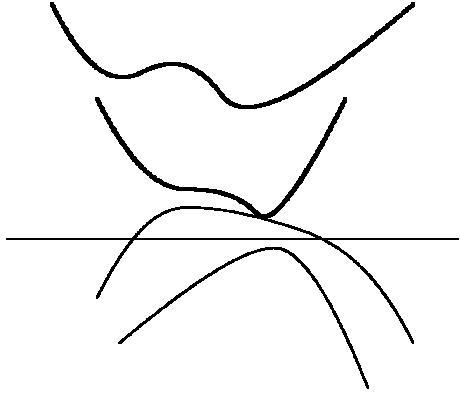


\begin{center}

\fignogap

\end{center}

\caption{
Possible configuration of 
singular loci when there is no $k_s^1$ contour avoiding singularity.} 
\label{fignogap}

\end{figure}
\begin{enumerate}
\item The closure of the
gap requires the collision of a pair of points on the singular loci, one 
on a thick line and the other on a thin line as in Fig.~\ref{fignogap}.
\item For the singular loci coming from the zeroes of $\ell_r^2+m_r^2$, the $\Imm(\ell^1_r)$
values on the loci for given $\Rea(k_s^1)$, determined by \refb{ecr22}, 
do not change during the second deformation since the $k_s^0$ and $k_{s\perp}$ contours
remain unchanged during this deformation. However the relation
between $\Imm(\ell_r^1)$ and $\Imm(k_s^1)$ changes, causing the vertical shift of the
singular loci in the $k_s^1$ plane as described above.
\item Since at the beginning of the second deformation $\Imm (\ell_r^1)$ was positive on the
thick curves and negative on the thin curves, they will continue to obey this, although the
sign of $\Imm(k_s^1)$ can change, as for example in going from Fig.~\ref{fignewrep} 
to Fig.~\ref{fignewrepmod}.
\end{enumerate}
Combining these observations we see that when the gap closes, one of the
propagators in the set $\BBB_s$ 
that go on-shell at the singular point has $\Imm(\ell_r^1)> 0$ and 
another on-shell propagator in this set has $\Imm(\ell_r^1)< 0$. 
Note the strict inequality: if we had $\Imm(\ell_r^1)=0$ then the corresponding
singular locus would have touched the real $k_s^1$ axis at the beginning of the second
deformation. This would have been inconsistent with the already proven result that at the
beginning of second deformation we can integrate $k_s^1$ along the real axis without
encountering any singularity.
In special cases there may be more 
propagators that become on-shell at the same point in the $k_s^1$ plane, 
causing more than two curves to touch at a point,
but none of these will have $\Imm(\ell_r^1)=0$ and at least a pair of them will have to carry opposite signs
of $\Imm(\ell_r^1)$.

So far we have been considering the possibility of avoiding singularities by deforming the
$k_s^1$ integration contour for one particular loop. Once that fails due to closure of the
gap, we still have the possibility of opening the gap by deforming the $k_{s'}^1$ integration
contour
for some other loop. Indeed such deformation of a neighboring loop momentum
will inject additional imaginary momentum along 1 direction through the vertices and
will change the imaginary parts of one or more $L_r^1$'s appearing in \refb{ecrdefLr}.  
This will induce further vertical shift of some of the singular loci in the $k_s^1$ plane, 
and could open up a
gap in Fig.~\ref{fignogap} 
through which the $k_s^1$ integration contour can pass. At the same time we have to
ensure that the deformation of the $k_{s'}^1$ integration contour does not run into a 
singularity on the way from some other propagator through which $k_{s'}$ flows.

\begin{figure}
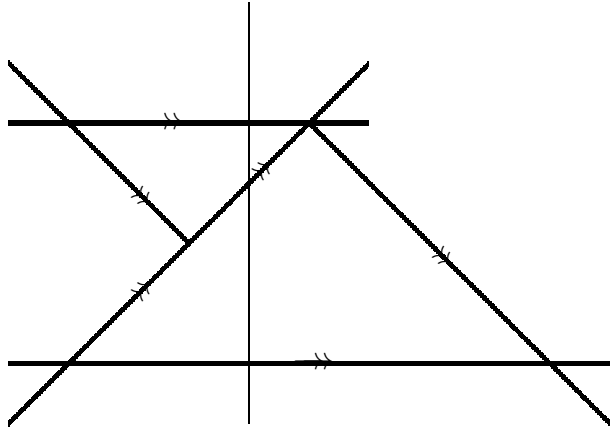


\begin{center}

\figarrowred

\end{center}

\caption{A reduced diagram labelling the direction of flow of $\Imm(\ell_i^1)$.
The flow is along the double arrows and always from left to right. \label{figarrowred}}

\end{figure}

After taking into account all of these possibilities we see that
the integral will run into a singularity
if we cannot deform the $\{k_s^1\}$ contours associated with 
different loops to avoid the singular
loci in some of the $k_s^1$ integrals.\footnote{This is a necessary but 
not sufficient condition
for running into singularities since we could still explore the possibility of making the 
$k_s^0$ integration contours depend on both real and imaginary components of $k_s^1$
and deforming the $k_s^i$ contours for $i\ge 2$ away from the real axis. We shall not
consider these possibilities.} We shall prove that such a situation does not arise as long
as \refb{ecr1new} holds. To prove this we assume the contrary, i.e.\ assume that we run
into a singularity and show that there is a contradiction. At the singularity certain set
of propagators become on-shell. We associate with the singularity a reduced diagram 
by contracting to points all propagators that are not on-shell, and associate to each 
on-shell propagator
an arrow that flows along $\ell_i$ if $\Imm(\ell_i^1)>0$ and opposite to $\ell_i$ if
$\Imm(\ell_i^1)<0$ at the singular point. 
Note that these arrows label different attributes compared to the arrows used in 
\S\ref{s2.1} -- to make this distinction we shall use double arrows for keeping track of the sign of
$\Imm(\ell_i^1)$.
Our earlier observation will now tell us that
along each loop there should be at least one propagator carrying an arrow along the
loop momentum and another propagator carrying the arrow opposite to the loop
momentum. Therefore we cannot have an oriented closed loop in the reduced diagram.
An argument identical to the one given in \S\ref{s2.1} now shows that we can 
partially order the vertices of the reduced diagram so that $\Imm(\ell^1)$ always
flows from the left to the right. An example of a reduced diagram of this type can be
found in Fig.~\ref{figarrowred}.
If we now draw a vertical line through the propagators 
that divides the diagram into two parts, as shown by the thin vertical line in Fig.~\ref{figarrowred},
then across the vertical line there will be on-shell
particles carrying momenta $\ell_i$, with 
$\Imm(\ell_i^1)$ flowing from left to the right.
We shall now show that this implies $\ell_{i\parallel I}\equiv\Imm(\ell_i^0, \ell_i^1)\in W^+$
and contradicts \refb{ecr1new}.

Let $\ell$ be the momentum carried
by an internal propagator and let $m$ be the mass of the particle propagating along that propagator. 
The pole coming from the denominator of this propagator is situated
at
\be
\ell^2 + m^2=0\, .
\ee
Writing $\ell = \ell_R + i \ell_I$ we get
\be 
0=\ell_R^2 - \ell_I^2 + 2 \,i \, \ell_R.\ell_I +m^2= \ell_{\parallel R}^2 + \ell_{\perp R}^2 - \ell_{\parallel I}^2 +
2 \,i \, \ell_{\parallel R}.\ell_{\parallel I}+m^2\, .
\ee
This gives
\be \label{ecr1.7}
\ell_{\parallel R}.\ell_{\parallel I}=0, \quad \ell_{\parallel R}^2 + \ell_{\perp R}^2 + m^2 - \ell_{\parallel I}^2=0\, .
\ee
Now it follows from the first equation that if $\ell_{\parallel R}\in V^\pm\cup (\p V^\pm-0)$ then 
$\ell_{\parallel I}\in 
W^\pm\cup \p W^\pm$ and
if $\ell_{\parallel R}\in W^\pm\cup (\p W^\pm-0)$ then $\ell_{\parallel I}\in V^\pm\cup\p V^\pm$.
Since vectors in $V^\pm$ have negative norm and
vectors in $W^\pm$ have positive norm, it follows that $\ell_{\parallel R}^2$ and $\ell_{\parallel I}^2$ have
opposite signs. Since $\ell_{\perp R}^2\ge 0$, it follows from the second equation in
\refb{ecr1.7} that $\ell_{\parallel R}^2\le 0$ and 
$\ell_{\parallel I}^2\ge 0$. If $\ell_{\parallel R}^2= 0$ but $\ell_{\parallel R}\ne 0$, then it follows from the first equation
of \refb{ecr1.7}  that we must also have 
$\ell_{\parallel I}^2 =0$, but this will be inconsistent with the second equation. Similar
contradiction arises if $\ell_{\parallel I}^2= 0$ but $\ell_{\parallel I}\ne 0$.
Therefore we must have $\ell_{\parallel R}^2 <0$ or $\ell_{\parallel R}=0$ and $\ell_{\parallel I}^2>0$ or 
$\ell_{\parallel I}=0$, but both $\ell_{\parallel R}$ and $\ell_{\parallel I}$ cannot vanish at the same time. This
may be summarized as
\be \label{ecrx1}
\ell_{\parallel R}\in V^\pm\cup 0, \quad \ell_{\parallel I}\in W^\pm\cup 0, \quad (\ell_{\parallel R},\ell_{\parallel I})
\ne (0,0) \, .
\ee

Consider now a vertical cut of a reduced diagram of the type shown in Fig.~\ref{figarrowred}. 
Let $\ell_i$ denote the momenta of the cut propagators, flowing
to the right. Since according to our previous analysis $\ell_{iI}^1> 0$, 
it follows from \refb{ecrx1} that 
$\ell_{i\parallel I}\in W^+$. Therefore
\be\label{ecrlili}
\sum_i \ell_{i\parallel I}\in W^+\, ,
\ee
where the sum runs over all the propagators cut by the vertical line. 
However, by momentum conservation we have
$\sum_i \ell_i = P_{(\alpha)}$ for some $P_{(\alpha)}$ describing the total external momenta entering the diagram from
the left. Therefore we have
\be 
P_{(\alpha)\parallel I}\in W^+\, .
\ee
This is inconsistent with \refb{ecr1new}.  Therefore we see that our initial assumption must
be wrong, namely that there is no singularity of the integral when \refb{ecr1new} is
satisfied.

Special attention may be paid to the boundary points $P_{(\alpha)}\in \p V^+$. In order to have a pinch
singularity on the integration contour, 
the conditions $\sum_i \ell_i = P_{(\alpha)}$ and \refb{ecrlili} will be compatible only if
$\ell_{i\parallel I}\in \p W^+$. Now we have already ruled out $\ell_{i\parallel I}=0$ since that would 
cause a singularity at the end of the first deformation. The first 
condition in \refb{ecr1.7}  now tells us that $\ell_{i\parallel R}\in \p V^\pm$. Therefore both $\ell_{i\parallel I}^2$ and
$\ell_{i\parallel R}^2$ vanishes. This would violate the second equation in \refb{ecr1.7} by finite amount. 
This shows that even for $P_{(\alpha)}\in (\p V^+-0)$, the contours are at finite distance away from
the pinch singularities. Therefore these points are in the interior of the domain of holomorphy, since
small deformations of external momenta will not produce a singularity.

\section{Discussion}

We shall end the paper by making a few comments on our results:
\begin{enumerate}
\item We have shown that when we define the Green's function at a given point in the complex momentum
space satisfying \refb{ecr1new} by analytic continuation of the Green's function at the origin along a particular path
lying inside the region \refb{ecr1new}, the result does not have have any singularity. One could however ask whether
the result could depend on the path along which we carry out the analytic continuation. To this end we note that
since the region \refb{ecr1new} is a simply connected region containing the origin (a star shaped region according to the
notation of \cite{BEG1,BEG2}) analytic continuation along any path lying wholly within the region \refb{ecr1new} will
always give the same result. 
\item We have proved a limited version of the general result obeyed by a local quantum field theory by restricting,
in \refb{ecr1new}, the imaginary part of the external momenta to lie in a two dimensional Lorentzian plane. This is
sufficient for the proof of crossing symmetry\cite{BEG2} and various other results on the
analyticity of the S-matrix\cite{itzykson}, but it will be of interest to extend the results to general external momenta
obeying only the condition \refb{ecrth}.

The arguments given in \S\ref{s2.2} can be used to proceed some way 
towards a proof of this more general result. Indeed since
at the end of \S\ref{s2.2} we established the existence of deformed integration contours avoiding poles when 
$p_a^0$ and $p_a^1$ have imaginary components, we can take this as the starting point and switch on imaginary
values of $p_a^2$. Proceeding as in \S\ref{s2.2} we can establish that we can associate a triple arrow to an internal 
propagator carrying momentum $\ell$ of
any reduced diagram that denotes the direction of $\Imm(\ell^2)$ at the supposed pinch singularity, 
and that these arrows never form a closed 
loop.\footnote{Note that $\ell^2$ in this discussion denotes the second component of $\ell$ and not the invariant
square of $\ell$.}
Therefore we can draw the diagrams so that $\Imm(\ell^2)$ flows from left to right and if we cut the diagram by
a vertical line then all the cut propagators carry $\Imm(\ell_i^2)$ from left to right. 
We also know from the on-shell condition that $\Imm(\ell_i)$ is space-like.
However unlike in the case
of \S\ref{s2.2} we cannot now conclude that the sum of the momenta carried by
these cut propagators have space-like imaginary part
-- this requires also the knowledge of $\Imm(\ell_i^1)$ carried by each cut propagator. The signs of the 
latter are labelled by the
double arrows which are not necessarily correlated with the triple arrows and can flow in either direction through
the cut propagators in the present case.

\item We have invoked the analysis given in \cite{BEG2} to conclude that the analyticity of 
the off-shell Green's function
in the domain \refb{ecr1new} is sufficient to prove crossing symmetry. This requires us to make use of some
general results on functions of many complex variables that states that the domain of analyticity of such functions
cannot have arbitrary shapes. Therefore analyticity in the domain \refb{ecr1new} actually implies analyticity in a larger
domain. Similar arguments can be used to prove various other analyticity properties of the S-matrix given the
analyticity of the Green's function in the domain \refb{ecr1new}\cite{itzykson}.
It will be interesting to explore if the momentum space analysis could directly establish analyticity
in the larger domain needed for the proof of crossing symmetry and other analyticity properties of the
S-matrix. This may then enable us to 
bypass the need of using an off-shell
formalism for the proof of crossing symmetry.
\end{enumerate}

\bigskip

{\bf Acknowledgement:}
We wish to thank Roji Pius and George Sterman for useful discussions.
The work of H.E.\ is conducted under a Carl Friedrich von Siemens Research Fellowship of the Alexander von Humboldt Foundation for postdoctoral researchers.
The work of A.S. was
supported in part by the J. C. Bose fellowship of 
the Department of Science and Technology, India and also by the Infosys Chair Professorship.

\end{document}